\documentclass[12pt]{iopart}

\usepackage[utf8]{inputenc}
\usepackage[headsep=15mm]{geometry}
\usepackage{amstext} 
\usepackage{iopams} 
\usepackage{hyperref}
\usepackage{xcolor}
\usepackage{graphicx}

\renewcommand\e[1]{\ensuremath{_{\text{#1}}}}

\usepackage{fancyhdr}
\fancypagestyle{plain}{
\fancyhf{}

\fancyhead[C]{\textsf{\href{http://dx.doi.org/10.1088/0953-8984/28/27/275201}{\textcolor{blue}{Published as: \emph{J. Phys. Condens. Matter} \textbf{2016}, 28, 275201,\\ DOI: 10.1088/0953-8984/28/27/275201}}}}
}
\begin{document}

\title[ELATE: analysis and visualization of elastic tensors]{ELATE: An open-source online application for analysis and visualization of elastic tensors}

\author{Romain Gaillac}
\address{Chimie ParisTech, PSL Research University, CNRS, Institut de Recherche de Chimie Paris, 75005 Paris, France}
\address{Air Liquide, Centre de Recherche Paris Saclay, 78354 Jouy-en-Josas, France}
\author{Pluton Pullumbi}
\address{Air Liquide, Centre de Recherche Paris Saclay, 78354 Jouy-en-Josas, France}
\author{Fran\c{c}ois-Xavier Coudert}
\address{Chimie ParisTech, PSL Research University, CNRS, Institut de Recherche de Chimie Paris, 75005 Paris, France}
\ead{fx.coudert@chimie-paristech.fr}

\begin{abstract}
We report on the implementation of a tool for the analysis of second-order elastic stiffness tensors, provided with both an open-source Python module and a standalone online application providing visualization tools of anisotropic mechanical properties. After describing the software features, how we compute the conventional elastic constants and how we represent them graphically, we explain our technical choices for the implementation. In particular, we focus on why a Python module is used to generate the HTML web page with embedded Javascript for dynamical plots.
\end{abstract}

\thispagestyle{plain}
\maketitle

\section{Introduction}

Despite the importance of the second-order elastic constants in the understanding of the mechanical properties of materials, they have been measured for a very small fraction of known crystalline materials. This is due to the need for large single crystals and the difficulty in precise experimental measurements, e.g. through Brillouin scattering,\cite{Vacher1972} in particular in low-symmetry crystals. This lack of experimental data limits the ability of materials scientists to develop new materials with targeted mechanical responses. On the other hand, the computational power available in high-performance computing (HPC) infrastructures has now reached the point where it is possible to make \emph{ab initio} predictions of elastic constants from crystalline structures by routine calculations.

Indeed, in the past few years, several quantum chemistry software packages have gained the capacity to calculate full second-order elastic tensors of crystals in an automated fashion. This can be implemented either internally, as part of the code of the quantum chemistry software package itself; or through a set of external wrapper scripts, which generate sets of input files, run them and analyze the output --- for example to derive energy--strain or stress--strain curves, from which the elastic constants are then determined. Popular computational chemistry tools for this purpose include:
\begin{itemize}
\item[--] \textsc{Crystal}\cite{CRYSTAL} gained this capability in its \textsc{Crystal09} version\cite{Perger2009,Erba2014}, and since version \textsc{Crystal14} can calculate piezoelectric and photoelastic tensors.\cite{Noel2001,Erba2013}
\item[--] VASP can calculate second-order elastic constants by strain-stress relationships since version 5.1.\cite{Lepage2002}
\item[--] ElaStic\footnote{\url{http://exciting-code.org/elastic}},\cite{Golesorkhtabar2013} a set of Python routines using DFT codes such as exciting, Wien2k, and Quantum Espresso as back ends.
\end{itemize}

Furthermore, this approach of \emph{ab initio} calculation of elastic constants has recently been taken to a larger scale, towards systematic mechanical characterization of known materials, by high-throughput calculations of elastic constants of inorganic materials as part of the Materials Project.\cite{Jain2013} de Jong et al. presented a very large database of calculated elastic properties for inorganic compounds, calculated at the density functional theory (DFT) level.\cite{deJong2015} The database contained elastic tensors for 1181 inorganic compounds upon publication in March 2015, and contains 2111 to date.

Given the fast-increasing availability of elastic constant databases, tools for their analysis and visualization are more important than ever. In particular, for anisotropic materials, there is a real need to analyze and visualize the directional elastic properties --- such as Young's modulus, linear compressibility, shear modulus and Poisson's ratio --- rather than their averages. Detailed tensorial analyses are also necessary to find materials with targeted or anomalous mechanical properties, such as negative linear compressibility (NLC),\cite{Cairns2015} negative Poisson's ratio (partial or total auxeticity),\cite{Evans1991,Siddorn2015} or highly-anisotropic elastic moduli.\cite{Ortiz_PRL2012} Although this tensorial analysis is well documented in textbooks and not fundamentally difficult, it is somewhat technical and casual users are not willing to invest time into developing their own implementation.

There exist, to our knowledge, two software packages currently available for the analysis of second-order elastic tensor and the visualization of elastic properties. The first one available is the code by Marmier, \textsc{ElAM}, in addition to a detailed, yet accessible, technical paper describing the transformations of the elastic tensor.\cite{Marmier} \textsc{ElAm}, implemented in Fortran 90, is command-line driven and can output 2D figures in PostScript format and 3D surfaces in the VRML (virtual reality modelling language) format. The use of this software requires compilation of the source code or local installation by the user does not allow interactive manipulation of the resulting graphs. In our own group's previous work,\cite{Ortiz2013} we relied upon an independent Mathematica implementation of tensorial analysis and visualization (whose source code we published as supporting information in Ref.~\cite{Ortiz2013}), which allows interactive use but is restricted to the proprietary Mathematica software suite.\cite{Mathematica}

Here, we describe ELATE, which is both an open source Python module (available at \url{https://github.com/fxcoudert/elate}) for the manipulation of elastic tensors and a standalone online application (available at \url{http://progs.coudert.name/elate}) for routine analysis of elastic tensors. In addition to being a user-friendly interface to the analysis code with no local installation required, the online application also provides a simple application programming interface (API) to be called from other materials-oriented web-based tools. Notably, it can also import elastic data directly from the Materials Project database of elastic constants by use of the Materials API (MAPI).

\section{Software description and features}

\subsection{Analysis and visualization of elastic properties}

As input, ELATE takes a $6\times 6$ symmetric matrix of second-order elastic constants, $C_{ij}$, in Voigt notation and units of GPa. It also takes, optionally, a user-defined system name for display purposes. From this starting point, the program first computes and display the main mechanical properties --- bulk modulus, Young's modulus, shear modulus and Poisson's ratio --- according to the three averaging schemes (Voigt,\cite{Voigt} Reuss\cite{Reuss} and Hill\cite{Hill}). The six eigenvalues of the elastic tensor are also computed. If at least one of them is negative, the elastic tensor is not definite positive, indicating mechanical instability of the system:\cite{Mouhat2014} the calculations stop there with an error message.

\begin{figure}[!ht]
\centering \includegraphics[width=0.9\textwidth]{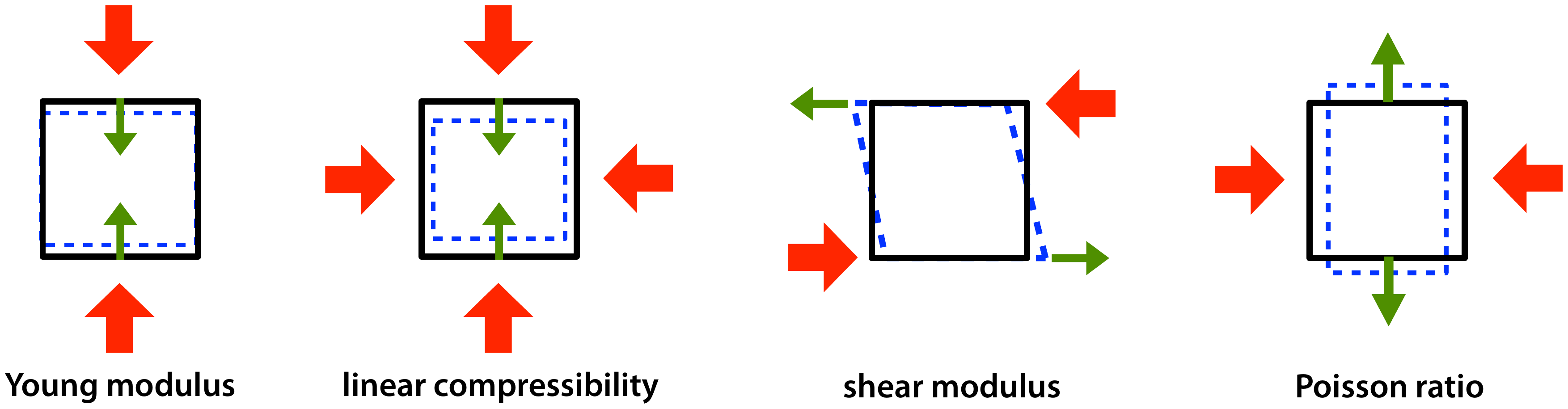}
\caption{Schematic representation of the directional elastic properties of materials: Young's modulus $E$, linear compressibility $\beta$, shear modulus $G$, and Poisson's ratio $\nu$. Red arrows represent the direction of stress exerted, green arrows the axis along which the response is measured.}
\label{fig:elasticity}
\end{figure}

\begin{figure}[!ht]
\centering \includegraphics[width=0.8\textwidth]{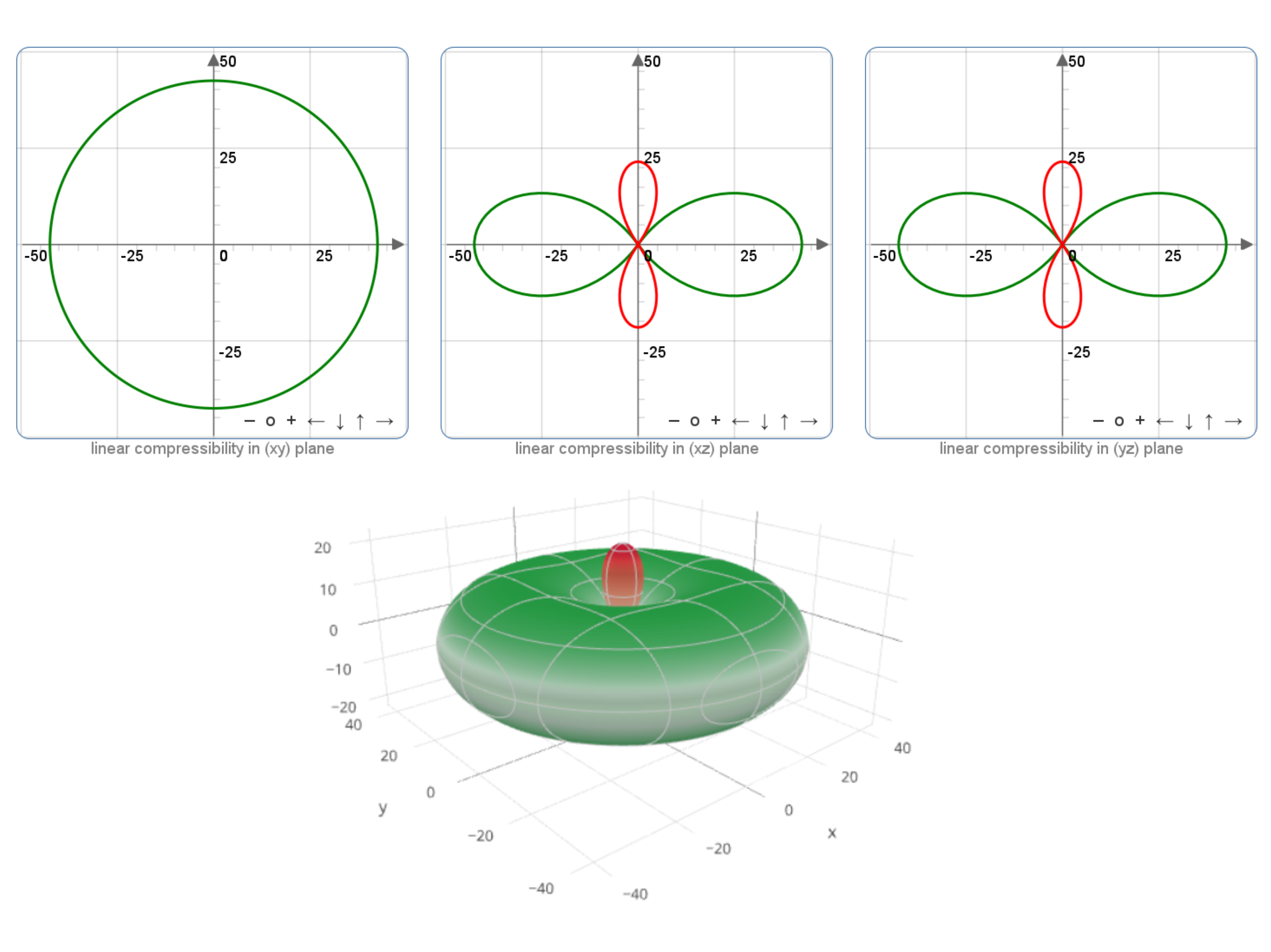}
\caption{Spatial dependence of the linear compressibility of Ag\e{3}Co(CN)\e{6}, a material in the trigonal crystal system which exhibits very large negative linear compressibility.\cite{Goodwin2008}. Elastic tensor was taken from the calculations of Ref.~\cite{Fang2014}.}
\label{fig:NLC}
\end{figure}

For mechanically stable systems, in addition to the average quantities, the program computes and displays additional information on the spatial variation of the elastic moduli (depicted in Figure~\ref{fig:elasticity}): Young's modulus ($E$), linear compressibility ($\beta$), shear modulus ($G$) and Poisson's ratio ($\nu$). The first two quantities are function of a single unit vector $\mathbf a$: $E(\mathbf a)$ and $\beta(\mathbf a)$, and can alternatively be parameterized by two angles in spherical coordinates: $0 \leq \theta \leq \pi$, and $0 \leq \varphi \leq 2\pi$:
\begin{equation}
\mathbf a = \left(
\begin{array}{c c c}
\sin(\theta)\cos(\varphi)\\
\sin(\theta)\sin(\varphi)\\
\cos(\theta)
\end{array}
\right)
\end{equation}
Both the Young's modulus and linear compressibility can thus be represented as a simple 3D parametric surface. ELATE allows direct visualization of the 3D spherical plot, as well as 2D projections on the $(xy)$, $(xz)$ and $(yz)$ planes. For linear compressibility (LC), which can be negative in some directions, two sets of surfaces (in 3D) and curves (in 2D) are plotted: directions corresponding to positive values of LC are plotted in green, and those of negative LC are plotted in red. This is exemplified on Figure~\ref{fig:NLC}, showing the linear compressibility of Ag\e{3}Co(CN)\e{6}.\cite{Goodwin2008, Fang2014}

The shear modulus $G$ and Poisson's ratio $\nu$ are not as straightforward to represent, because they depend on two orthogonal unit vectors $\mathbf a$ and $\mathbf b$ (respectively the direction of the stress applied and the direction of measurement). Here again, those two unit vectors can be parameterized by three angles, so that $\mathbf b$ is written as:\footnote{We note here that there is a typo in the expression of $\mathbf b$ in Eq.~6 of Ref.~\cite{Marmier}.}
\begin{equation}
	\mathbf b = \left(
	\begin{array}{c c c}
	\cos(\theta)\cos(\varphi)\cos(\chi)-\sin(\phi)\sin(\chi)\\
	\cos(\theta)\sin(\varphi)\cos(\chi)+\cos(\phi)\sin(\chi)\\
	-\sin(\theta)\cos(\chi)
	\end{array}
	\right)
\end{equation}
with $0 \leq \chi \leq 2\pi$. Since $G(\theta, \varphi, \chi)$ and $\nu(\theta, \varphi, \chi)$ cannot be directly represented in 3D space, we followed Marmier\cite{Marmier} in their representation, plotting each as spherical $(\theta, \varphi)$ coordinates the minimal and maximal values over all possible values of $\chi$: $f(\theta, \varphi) = \min_\chi X(\theta, \varphi, \chi)$ and $g(\theta, \varphi) = \max_\chi X(\theta, \varphi, \chi)$, respectively. The surface representing $g$ encloses that representing $f$, so it is plotted in translucent blue color. $f$ is represented as an surface, with solid green lobes for positive values and translucent red lobes for negative values (for the Poisson's ratio, which can be negative in some directions).

\begin{figure}[!ht]
\centering \includegraphics[width=0.8\textwidth]{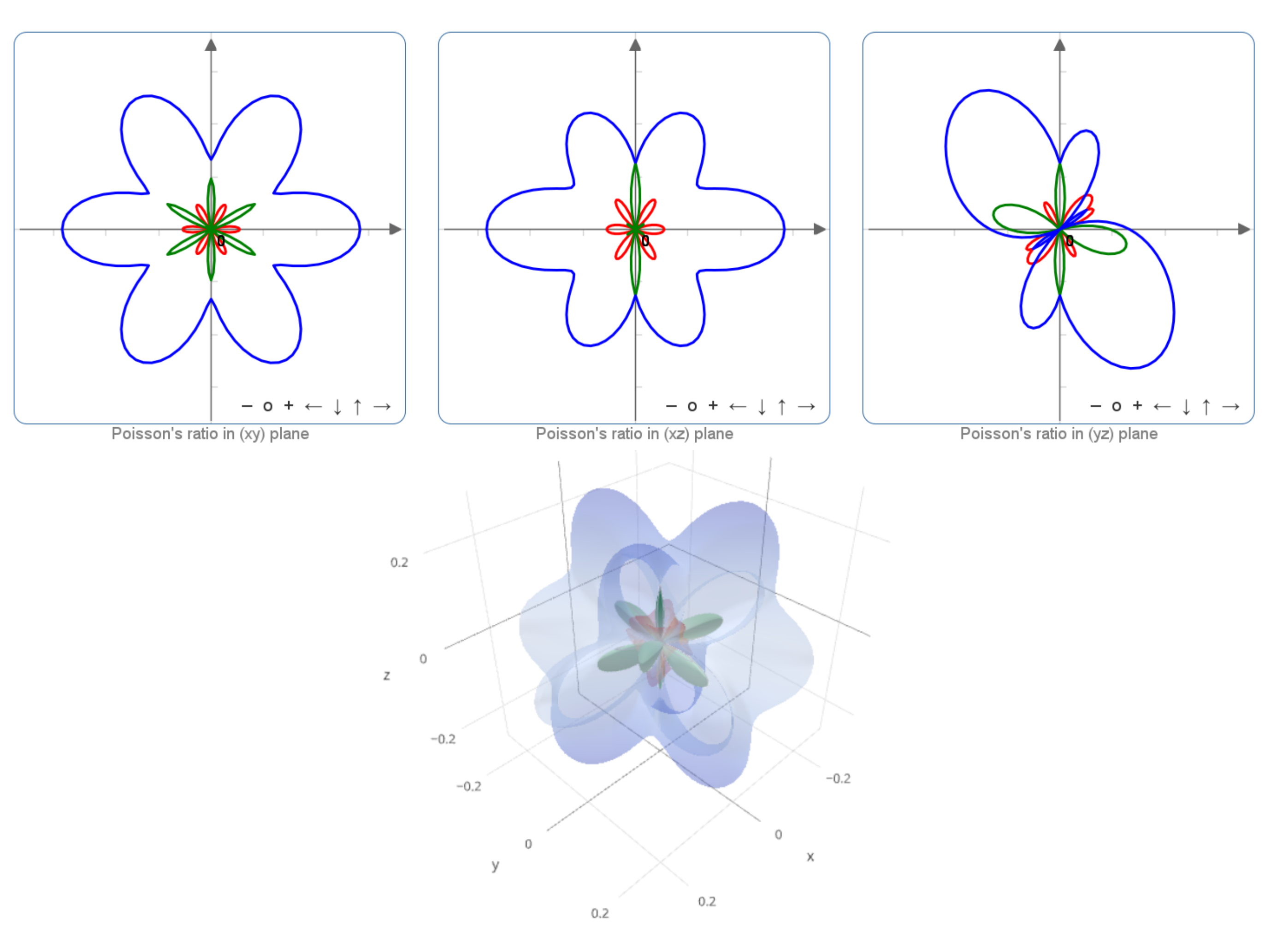}
\caption{Spatial dependence of the Poisson's ratio of $\alpha$-quartz (trigonal crystal system).}
\label{fig:Poisson_NSI}
\end{figure}

Figure~\ref{fig:Poisson_NSI} shows, on the example of $\alpha$-quartz, the three two-dimensional plots and the three-dimensional plot all representing the Poisson's ratio of this material.

In addition to the 2D and 3D graphical representations of the directional elastic properties above, ELATE also provides a quantitative analysis by reporting the minimal and maximal values of each modulus (e.g., $E\e{min}$ and $E\e{max}$), as well as the directions ($\mathbf{a}\e{min/max}$ and $\mathbf{b}\e{min/max}$) along which these extrema occur. This allows the determination of directions of particular interest in the elastic properties, which are not necessarily along the crystallographic axes of the material. We also report a measure of the anisotropy $A_X$ of each elastic modulus $X$, defined as follows:
\begin{equation}
A_{X} =
\left\lbrace
\begin{array}{l l l}
X_{max} / X_{min} \text{ if } \textrm{sign}(X_{max}) = \textrm{sign}(X_{min})  \\[2mm]
\infty \text{ otherwise}
\end{array}
\right.
\end{equation}
This measure is particularly interesting as marked anisotropy of the mechanical properties is often associated with anomalous mechanical behavior,\cite{Coudert2015} including the possibility for large negative linear compressibility\cite{Cairns2015} or negative area compressibility,\cite{Ogborn2012} or the potential for adsorption- or pressure-induced large-scale structural transitions.\cite{Ortiz_PRL2012}

\subsection{Technical choices and implementation}

Regarding the theoretical derivations of the mechanical properties depending on the elastic constants, we essentially refer the reader to excellent paper by Marmier,\cite{Marmier} upon which our implementation of the tensorial analysis is based. We provide here some details about the technical implementation of the online ELATE application, and its interface with the Materials Project database.

In terms of implementation, we opted for a client-server architecture where the computations are performed directly on the server through Python code. The most computationally expensive part of the tensorial analysis consists in calculating the values for plots of the directional shear modulus and Poisson's ratio, since they rely on numerical minimization along the $\chi$ coordinate. Those would have been both difficult to implement and slow to run in the browser, for example in JavaScript. On the other hand, the Python \texttt{scipy} and \texttt{numpy} libraries provide great tools for linear algebra, needed for the computation itself, and minimization algorithms, needed in order to plot the results. Thus the server performs the calculation of values necessary for 2D and 3D plots, embeds them into the resulting HTML page along with the JavaScript code for rendering them. The graphs are then displayed in the browser, and the user can interact with them without any lag: this includes zooming, rotating, panning, as well as inspecting values at points of interest.

The 2D graphs are displayed using the JSXGraph cross-browser library\cite{Gerhauser2011,jsxgraph}, developed at the \emph{Lehrstuhl f\"{u}r Mathematik und ihre Didaktik} of the University of Bayreuth. Visualization of 3D surfaces is done using \texttt{plotly.js}\cite{plotly}, a free and recently open-sourced graphing library.\cite{plotly_opensource} With that framework, we are able to represent dynamic parametric surfaces in a browser, making the spatial representation of mechanical properties much easier and fully interactive. The web content is served by Bottle, a very simple Python Web framework.\cite{bottle}

\subsection{Interfacing with the Materials Project database}

In addition to the possibility for the user to input a second-order elastic tensor directly into the application, we provide a layer of integration with the elastic data available from the Materials Project (\cite{Jain2013,deJong2015}). This database contain elastic tensors for more then 2,000 crystalline inorganic materials, obtained by quantum chemical calculations at the Density Functional Theory (DFT) level. By interfacing our web application with the Materials API, one can search for any material in the database in three different ways: (i) by providing a Materials Project identifier (e.g., \texttt{mp-2133}), (ii) by looking up a given chemical formula (such as ZnO), or (iii) by searching for a combination of elements (such as Li-Fe-O). When the request returns several results, the user is prompted with at most ten representatives of the chemical system he was looking for, indicating which one has elastic data available. The elastic properties can then be directly analyzed and visualized in ELATE, including a link to the relevant page of the Materials Project database to browser the compound's properties.

We also provide a very simple API to allow other online applications, such as the Materials Project database browser, to link to ELATE visualizations: for any Materials Project identifier \texttt{\emph{id}}, accessing the URL \texttt{http://progs.coudert.name/elate/mp?\emph{id}} will directly open on the detailed analysis of the elastic properties of the material. Integration to other databases could be provided in the future, as the needs of the community evolve.

\section{Perspectives}

As computational power increases in time, databases of computational properties of materials, both experimentally known and hypothetical, grow larger in size and contain more properties, going beyond structures and energies of formation. The availability of this data requires tools to analyze it and visualize these properties, in order to be able to identify best performers for targeted properties, detect trends and understand relationships between them. We provide here ELATE, a Python module and an online application for the analysis and visualization of second-order elastic tensors, providing some integration with the Materials Project database. In the future, we look forward to interface it with other available databases of mechanical properties, hopefully constructing a single and coherent Application Programming Interface (API) rather than coexisting incompatible protocols. As additional categories of properties become available through this large-scale computational approach, we also intend to extend ELATE to analyze other tensorial properties, such as piezoelectric and photoelastic tensors.

\bigskip\bigskip
\section*{Acknowledgments}

We thank Jack Evans for $\beta$-testing ELATE and providing useful feedback. This work benefitted from the support of ANRT (th\`ese CIFRE 2015/0268).

\section*{References}

\bibliographystyle{iopart-num}
\bibliography{biblio}

\providecommand{\newblock}{}
\begin{thebibliography}{10}
\expandafter\ifx\csname url\endcsname\relax
  \def\url#1{{\tt #1}}\fi
\expandafter\ifx\csname urlprefix\endcsname\relax\def\urlprefix{URL }\fi
\providecommand{\eprint}[2][]{\url{#2}}

\bibitem{Vacher1972}
Vacher R and Boyer L 1972 {\em Phys. Rev. B\/} {\bf 6} 639--673

\bibitem{CRYSTAL}
Dovesi R, Orlando R, Civalleri B, Roetti C, Saunders V~R and Zicovich-Wilson
  C~M 2005 {\em Z. Kristallogr.\/} {\bf 220} 571--573

\bibitem{Perger2009}
Perger W, Criswell J, Civalleri B and Dovesi R 2009 {\em Comput. Phys.
  Commun.\/} {\bf 180} 1753--1759

\bibitem{Erba2014}
Erba A, Mahmoud A, Orlando R and Dovesi R 2014 {\em Phys Chem Minerals\/} {\bf
  41} 161--162

\bibitem{Noel2001}
Noel Y, Zicovich-Wilson C~M, Civalleri B, D'Arco P and Dovesi R 2001 {\em Phys.
  Rev. B\/} {\bf 65}

\bibitem{Erba2013}
Erba A, El-Kelany K~E, Ferrero M, Baraille I and R\'erat M 2013 {\em Phys. Rev.
  B\/} {\bf 88}

\bibitem{Lepage2002}
Le~Page Y and Saxe P 2002 {\em Phys. Rev. B\/} {\bf 65}

\bibitem{Golesorkhtabar2013}
Golesorkhtabar R, Pavone P, Spitaler J, Puschnig P and Draxl C 2013 {\em
  Computer Physics Communications\/} {\bf 184} 1861--1873

\bibitem{Jain2013}
Jain A, Ong S~P, Hautier G, Chen W, Richards W~D, Dacek S, Cholia S, Gunter D,
  Skinner D, Ceder G and Persson K~A 2013 {\em APL Mater.\/} {\bf 1} 011002

\bibitem{deJong2015}
de~Jong M, Chen W, Angsten T, Jain A, Notestine R, Gamst A, Sluiter M,
  Krishna~Ande C, van~der Zwaag S, Plata J~J, Toher C, Curtarolo S, Ceder G,
  Persson K~A and Asta M 2015 {\em Sci. Data\/} {\bf 2} 150009

\bibitem{Cairns2015}
Cairns A~B and Goodwin A~L 2015 {\em Phys. Chem. Chem. Phys.\/} {\bf 17}
  20449--20465

\bibitem{Evans1991}
Evans K~E 1991 {\em Endeavour\/} {\bf 15} 170--174

\bibitem{Siddorn2015}
Siddorn M, Coudert F~X, Evans K~E and Marmier A 2015 {\em Phys. Chem. Chem.
  Phys.\/} {\bf 17} 17927--17933

\bibitem{Ortiz_PRL2012}
Ortiz A~U, Boutin A, Fuchs A~H and Coudert F~X 2012 {\em Phys. Rev. Lett.\/}
  {\bf 109} 195502

\bibitem{Marmier}
Marmier A, Lethbridge Z~A~D, Walton R~I, Smith C~W, Parker S~C and Evans K~E
  2010 {\em Comput. Phys. Commun.\/} {\bf 181} 2102--2115

\bibitem{Ortiz2013}
Ortiz A~U, Boutin A, Fuchs A~H and Coudert F~X 2013 {\em J. Chem. Phys.\/} {\bf
  138} 174703

\bibitem{Mathematica}
Wolfram Research, Inc., Mathematica, Version 10.3, Champaign, IL (2015).

\bibitem{Voigt}
Voigt W 1928 {\em Lehrbuch der Kristallphysik\/} (Teubner)

\bibitem{Reuss}
Reuss A 1929 {\em Z. angew. Math. Mech.\/} {\bf 9} 49--58

\bibitem{Hill}
Hill R 1952 {\em Proc. Phys. Soc. A\/} {\bf 65} 349--354

\bibitem{Mouhat2014}
Mouhat F and Coudert F~X 2014 {\em Phys. Rev. B\/} {\bf 90} 224104

\bibitem{Goodwin2008}
Goodwin A~L, Keen D~A and Tucker M~G 2008 {\em Proc. Nat. Acad. Sci.\/} {\bf
  105} 18708--18713

\bibitem{Fang2014}
Fang H, Dove M~T and Refson K 2014 {\em Physical Review B\/} {\bf 90}

\bibitem{Coudert2015}
Coudert F~X 2015 {\em Chem. Mater.\/} {\bf 27} 1905--1916

\bibitem{Ogborn2012}
Ogborn J~M, Collings I~E, Moggach S~A, Thompson A~L and Goodwin A~L 2012 {\em
  Chem. Sci.\/} {\bf 3} 3011

\bibitem{Gerhauser2011}
Gerh{\"a}user M, Miller C, Valentin B, Wassermann A and Wilfahrt P 2011 {\em
  Electronic Journal of Mathematics and Technology\/} {\bf 5}

\bibitem{jsxgraph}
JSXGraph, version 0.99.3, available online at
  \url{http://jsxgraph.uni-bayreuth.de/}

\bibitem{plotly}
Plotly Technologies Inc. Collaborative data science. Montréal, QC, 2015.
  Available online at \url{https://plot.ly}

\bibitem{plotly_opensource}
@plotlygraphs, Twitter, 17 November 2015,
  \url{https://twitter.com/plotlygraphs/status/666667595903459328}

\bibitem{bottle}
Bottle, version 0.13-dev dated 23 January 2016, available online at
  \url{http://bottlepy.org/}

\end{thebibliography}

\end{document}